\documentclass[10pt,letterpaper]{article}
\usepackage{opex3}

\begin{document}

\title{Realization of two Fourier-limited solid-state
single-photon sources}

\author{R. Lettow$^{1}$, V. Ahtee$^{1, 2}$, R. Pfab$^{1}$, A. Renn$^{1}$, E. Ikonen$^{2}$, S. G\"{o}tzinger$^{1}$, and V. Sandoghdar$^{1}$}

\address{$^{1}$Laboratory of Physical Chemistry and optETH, ETH Z\"{u}rich, CH-8093 Z\"{u}rich,
Switzerland}
\address{$^{2}$Metrology Research Institute, Helsinki University of Technology (TKK)
and Centre for Metrology and Accreditation (MIKES), P.O. Box 3000,
FI-02015 TKK, Finland}

\email{Stephan.goetzinger@phys.chem.ethz.ch} 

\homepage{http://www.nano-optics.ethz.ch}

\begin{abstract}
We demonstrate two solid-state sources of indistinguishable single
photons. High resolution laser spectroscopy and optical microscopy
were combined at $T=1.4~K$ to identify individual molecules in two
independent microscopes. The Stark effect was exploited to shift the
transition frequency of a given molecule and thus obtain single
photon sources with perfect spectral overlap. Our experimental
arrangement sets the ground for the realization of various quantum
interference and information processing experiments.
\end{abstract}




Single-photon sources have been demonstrated in the solid state with
molecules~\cite{Brunel:99,Lounis00}, quantum dots~\cite{Michler00}
and nitrogen-vacancy color centers in diamond~\cite{Kurtsiefer00} as
well as with atoms~\cite{Kuhn02} and ions~\cite{Keller:04} in the
gas phase. Single photons have been successfully used for
applications in quantum cryptography~\cite{Waks02,Alleaume04}, but
complex schemes of quantum information processing require a large
number of indistinguishable photons~\cite{Knill01,kok07}. The first
efforts in this direction have used consecutively emitted photons by
single quantum dots, atoms in cavities or
molecules~\cite{Santori02,Legero:04,Kiraz:05}. More recently
experiments with two independently trapped atoms or ions have been
also reported~\cite{Beugnon06,Maunz07}. Here, we show that two
molecules embedded in solid samples can also be used to generate
lifetime-limited photons for use in the investigation of photon
interference and correlation effects~\cite{Mandel:83}. The
solid-state aspect of our system offers many advantages including
well defined polarization and a nearly indefinite measurement time
using the same single emitter.

\begin{figure}
   \begin{center}
   \begin{tabular}{c}
   \includegraphics[height=4cm]{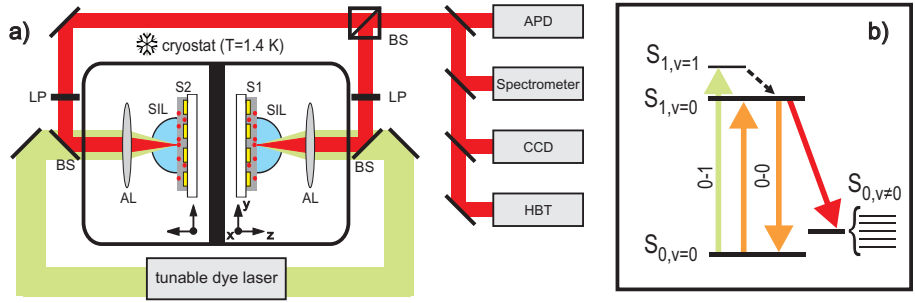}
   \end{tabular}
   \end{center}
   \caption[example]
   {\label{setup}
a) Experimental setup: Two low temperature microscopes with solid
immersion lenses are placed in a liquid helium bath cryostat. AL:
aspheric lens, SIL: solid-immersion lens, S: sample, LP: long pass
filter, BS: beam splitter, HBT: Hanbury Brown and Twiss
autocorrelator, consisting of a beam splitter, two APD´s and a time
correlator card (Pico Harp, Pico Quant). Details are given in the
text. b) Jablonsky diagram of a dye molecule with the relevant
energy levels.}
   \end{figure}

\label{sect:experi}

The experimental arrangement is sketched in Fig.~\ref{setup}~(a).
Two independent low temperature microscopes are separated by an
opaque wall in a liquid helium bath cryostat at $T=1.4$\,K. Light
from a tunable dye laser (Coherent 899-29, $\lambda=590$\, nm,
linewidth~$1$\, MHz) is coupled into the microscopes via galvo-optic
mirror scanners and telecentric lens systems. Each sample is
prepared by sandwiching a solution of dibenzanthanthrene (DBATT) in
\emph{n}-tetradecane (concentration of $10^{-7}/mol$) between a
glass substrate that contains interdigitating gold electrodes
(spacing $18$\, $\mu$m)~\cite{Kador:01} and a hemispherical cubic
zirconia solid immersion lens (SIL)~\cite{Mansfield:90}. The
electrodes can be used to apply electric fields of up to
$5\times10^{6}$\, V/m resulting in a Stark shift of more than $5$\,
GHz for the molecular resonance. The SILs combined with aspheric
lenses of high numerical aperture (NA=0.55) provide an effective NA
of $1.12$~\cite{wrigge}, allowing a high spatial resolution and an
enhanced fluorescence collection efficiency~\cite{koyama:99}. Three
piezoelectric slider systems in each microscope make it possible to
position the sample/SIL unit precisely with respect to the aspheric
lens. The collected photons of the microscopes can be sent either to
an avalanche photodiode (APD), a spectrometer with a resolution of
$0.3$~nm, an imaging CCD camera or a Hanbury Brown and Twiss (HBT)
type photon correlator. Long-pass cut-off and band-pass filters are
used to block undesired photons.

The energy spectrum of a dye molecule embedded in a solid at low
temperatures is composed of electronic and vibrational levels. In
what follows, we denote the states corresponding to the electronic
ground and the first excited states of DBATT by S$_{0,v}$ and
S$_{1,v}$, respectively (see fig.~1(b)). It turns out that each
vibrational transition is composed of a narrow zero-phonon line
(ZPL) and a phonon wing that stems from the vibrational coupling of
the molecule with the host matrix. The ratio of the photons emitted
into the ZPL to the total number of photons emitted into that
particular vibronic transition is given by the Debye-Waller
($\alpha_{DW}$). The ZPL of the transition between S$_{0,v=0}$ and
S$_{1,v=0}$ (from now on referred to as 0-0 ZPL) in DBATT becomes
natural linewidth limited at $T \le 2~K$, providing an ideal source
of Fourier-limited single photons.

\begin{figure} [h!]
   \begin{center}
   \begin{tabular}{c}
   \includegraphics[height=4cm]{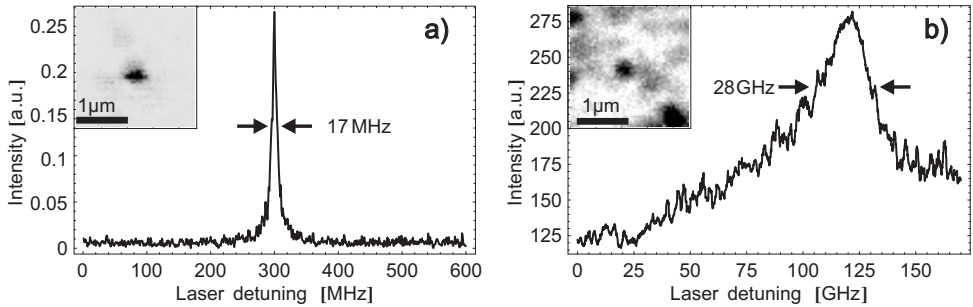}
   \end{tabular}
   \end{center}
   \caption[example]
   {\label{excitation-spec}
(a) Fluorescence excitation spectrum of a single molecule excited
via the 0-0 zero photon line. (b) Fluorescence excitation spectrum
of the same molecule excited via the 0-1 transition. Insets: Laser
scanning images of the molecule and its surrounding recorded at
T=$1.4$\, K under respective excitation conditions.}
   \end{figure}

In standard fluorescence excitation spectroscopy~\cite{Orrit90} a
narrow-band laser resonantly excites the molecule via the 0-0~ZPL.
The molecular excitation decays consequently via the vibrational
levels S$_{0, v}$, emitting Stokes-shifted photons (figure
\ref{setup} (b)). To achieve a high signal-to-noise ratio and a low
background in detecting single molecules, however, the 0-0 ZPL
emission is cut off by using filters in detection.
Figure~\ref{excitation-spec} (a) displays an excitation spectrum
from a single DBATT molecule. The full width at half-maximum (FWHM)
of 17$\pm1$ ~MHz corresponds to the excited state lifetime of
9.4$\pm0.5$ ns~\cite{boiron02}.

In order to capture the \emph{emission} on the narrow 0-0~ZPL
transition, we apply a second version of fluorescence excitation
spectroscopy by exciting the molecule to
S$_{1,v=1}$~\cite{nonn01,Kiraz:03} at $242$~cm$^{-1}$ higher
frequency than the 0-0~ZPL~\cite{jelezko02}. This state quickly
relaxes into S$_{1,v=0}$ where it decays further to the electronic
ground state and emits a visible photon as in the previous case
(figure \ref{setup}~(b)). Figure~\ref{excitation-spec} (b) shows
such an excitation spectrum for the same molecule studied in
Fig.~\ref{excitation-spec} (a). The observed FWHM of $28$~GHz is
about three orders of magnitude broader than the ZPL transition and
is caused by the short lifetime of about 5 ps for the S$_{1,v=1}$
state.

The large linewidth of the 0-1 transition implies its low absorption
cross section. As a result, more than two orders of magnitude higher
pump power ($100$\,nW) is needed to obtain a signal comparable to
the conventional excitation scheme used in
Fig.~\ref{excitation-spec} (a). This high laser intensity increases
the background fluorescence because together with broad resonance
lines it facilitates the excitation of neighboring molecules,
reducing the spectral selectivity that has been the central asset in
cryogenic single molecule detection. To get around this problem, we
have worked with samples that are 100 times more dilute than
commonly studied in single molecule spectroscopy. Furthermore, we
have exploited the high resolution provided by SILs to identify
single molecules spatially. The insets in
Figs.~\ref{excitation-spec} (a) and (b) show confocal scan images
recorded when the laser frequency was set to the resonances shown in
each figure. The FWHM of 290~nm is very close to the diffraction
limited value of 270~nm expected from NA=1.12 and allows us to
verify that the same molecule is the source of signals in both
cases.

\begin{figure} []
   \begin{center}
   \begin{tabular}{c}
   \includegraphics[height=5cm]{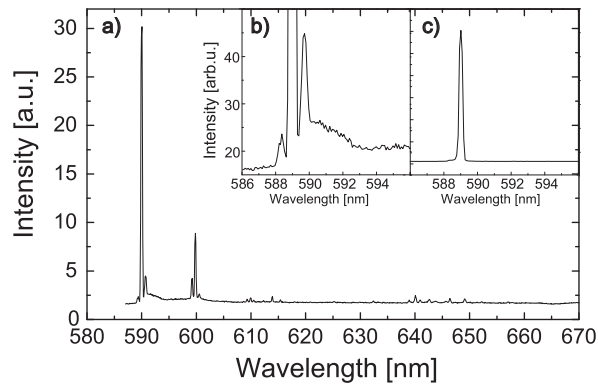}
   \end{tabular}
   \end{center}
   \caption[example]
   {\label{emission-spec}
a) Spectrum of a single DBATT molecule in \emph{n}-tetradecane under
0-1 excitation. The linewidth is limited by the spectrometer
resolution. b) Zoom around the 0-0 ZPL. The phonon wing, a local
mode on the longer wavelength side and a hot mode are clearly
visible. c) The 0-0 ZPL isolated by inserting a $0.5$~nm bandpass
filter. }
   \end{figure}

The next step toward the preparation of a Fourier-limited single
photon source is to filter the 0-0 phonon wing and the Stokes
shifted emission to extract the 0-0~ZPL. Figure~\ref{emission-spec}
(a) shows the emission spectrum of a single DBATT molecule obtained
by a 0-1 excitation. The dominant line at 590~nm represents the 0-0
ZPL. Several emission lines and the phonon wings (see also
Fig.~\ref{emission-spec} (b)) are clearly visible. The recorded
spectrum is in good agreement with ensemble measurements reported in
the literature~\cite{jelezko02}. By comparing the different areas
under the measured spectrum, we can estimate the Franck-Condon and
the Debye-Waller factors to be $0.4$ and $0.7$, respectively.
Figure~\ref{emission-spec} (c) shows the 0-0~ZPL after inserting a
$0.5$~nm bandpass filter in the detection path. In saturation
experiments we were able to measure more than $3.5 \times 10^{5}$
photons per second emitted by this line.

\begin{figure}
   \begin{center}
   \begin{tabular}{c}
   \includegraphics[height=4cm]{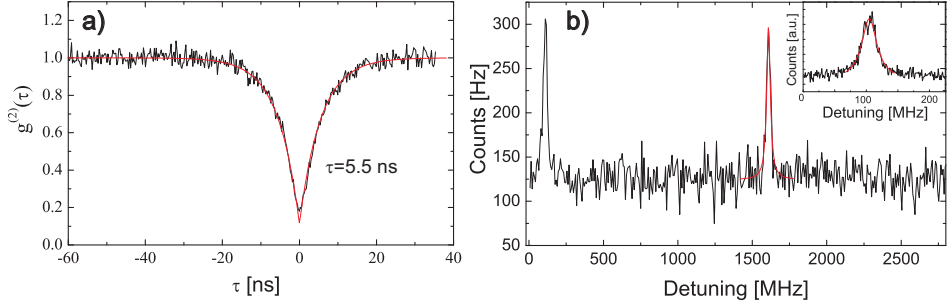}
   \end{tabular}
   \end{center}
   \caption[example]
   {\label{antibunch}
(a) Second order photon correlation measurement of the zero-phonon
line for a molecule excited via the 0-1 transition. Strong
antibunching is clearly visible. (b) Fluorescence of the 0-0 ZPL
analyzed by a Fabry-Perot cavity. Inset: A high resolution scan.}
   \end{figure}

To verify the realization of a single photon source, we have
measured the second order autocorrelation function $g^{(2)}(\tau)$
on the 0-0 ZPL isolated according to the above-mentioned procedure.
As shown in Fig.~\ref{antibunch}~(a), a strong antibunching is
observed with $g^{(2)}(0)=0.18$. We attribute the deviation from the
ideal value of $0$ to a limited time resolution of the single photon
counters. The antibunching characteristic time of $5.5$~ns is
shorter than the decay time of 9.4 ns for the S$_{1,v=0}$ state
because of the strong pumping of the population out of the ground
state~\cite{santori05}. To ensure that the emitted 0-0~ZPL photons
are nevertheless natural linewidth limited, we sent the emitted
photons through a Fabry-Perot cavity with a finesse of $200$ and a
free spectral range of 1.5~GHz. Figure~\ref{antibunch}~(b) clearly
shows a single peak within the free spectral range of the spectrum
analyzer. The zoom in the inset highlights the narrow linewidth when
accounting for the instrument FWHM of 7.5~MHz.

\begin{figure} [h!]
   \begin{center}
   \begin{tabular}{c}
   \includegraphics[height=5.5cm]{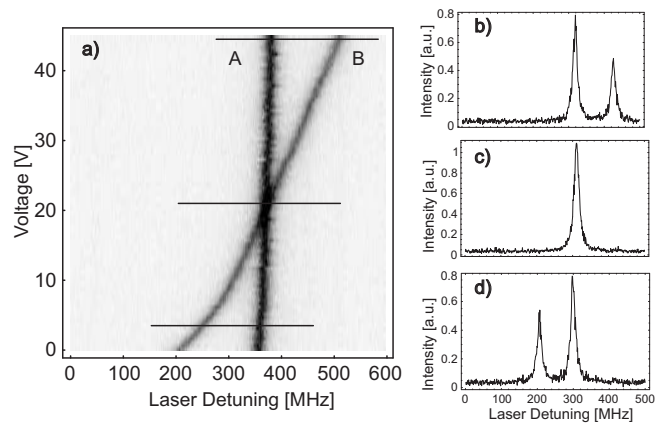}
   \end{tabular}
   \end{center}
   \caption[example]
   {\label{Stark}
(a) Fluorescence excitation spectra of two molecules. Molecule (B)
is shifted with respect to molecule (A) with increasing voltage.
(b)-(d) cross sections as indicated in (a). At $V=21$\,V, the two
spectra are fully overlapped.}
   \end{figure}

Having demonstrated the generation of narrow-band single photons
from single DBATT molecules, we searched for two molecules, one in
each microscope, that had 0-0 ZPLs within a few GHz of each other.
By applying a constant electric field to the electrodes embedded in
the glass substrates, we could tune the 0-0 ZPLs of the molecules
independently via the Stark effect. Figure~\ref{Stark} shows the 0-0
fluorescence excitation spectra from two molecules recorded on one
APD as the voltage applied to one sample was varied. The 0-0 ZPL of
one molecule shifts with increasing voltage and sweeps through the
resonance of the second molecule. At an applied field of about
$1.2\times10^{6}$~V/m the spectra of the two molecules can be no
longer distinguished. Given the well-defined orientation of the
transition dipole moment in DBATT, the emission of single molecules
can be linearly polarized with an extinction ratio of the order of
300:1~\cite{wrigge}, making it a simple task to obtain the same
polarization state for both single photon sources. Overlapping
Fourier-limited spectral lines and polarization states establish the
two molecules as sources of indistinguishable single photons.

In conclusion, we have combined cryogenic optical microscopy and
spectroscopy to identify and prepare two single molecules that
generate indistinguishable photons. A spectral analysis using a
Fabry-Perot cavity and antibunching measurements gave direct proof
for the generation of lifetime-limited single photons. The
solid-state arrangement in this experiment using SILs makes it
possible to miniaturize each single-photon source and to integrate
several of them on one chip. Together with the nearly indefinite
photostability and the long coherence times of dye molecules, these
features make our approach promising for performing complex quantum
operations. We are currently working to trigger the emission of
single photons by incorporating a pulsed laser for the 0-1
excitation.

\section{Acknowledgement}
This work was supported by the ETH Zurich via the INIT program
Quantum System for Information Technology (QSIT), the Swiss National
Science Foundation and the Academy of Finland (grant number 210857).
We thank G. Wrigge and I. Gerhardt for experimental help.

\end{document}